\documentclass[11pt]{article}

\usepackage{fullpage}
\usepackage{verbatim}
\usepackage{amsmath}
\usepackage{amsthm}
\usepackage{amsfonts}
\usepackage{epsfig}

\renewcommand{\Re}{\mathbb{R}}

\theoremstyle{plain}
 \newtheorem{lemma}{Lemma}[section]
 \newtheorem{theorem}[lemma]{Theorem}
 \newtheorem{corollary}[lemma]{Corollary}
 
 \newtheorem{proposition}[lemma]{Proposition}

 \theoremstyle{definition}
 \newtheorem{definition}[lemma]{Definition}
 
 \newtheorem{observation}[lemma]{Observation}

\newcommand{\rhst}{\textsc{Rhst}}
\newcommand{\combn}{\textsc{Combined}}
\newcommand{\wcombn}{\textsc{WCombined}}

\newenvironment{algrthree}{\medskip \noindent \textsc{Algorithm {\combn}\ }}{}
\newenvironment{combined}{\medskip \noindent \textsc{Algorithm $A$.\ }}{}
\newenvironment{RHST}{\medskip \noindent \textsc{Algorithm {\rhst}.\ }}{}

 \DeclareMathOperator{\dist}{d}
\DeclareMathOperator{\diam}{diam}
\DeclareMathOperator{\cost}{cost}
\DeclareMathOperator{\mcost}{\text{mcost`}}
\newcommand{\ie}{\emph{i.e.}}
\newcommand{\etal}{\emph{et. al.}}
\newcommand{\hst}{HST}
\newcommand{\aopt}{\ensuremath{\alpha\text{-}\mathrm{OPT}}}
\newcommand{\e}{\ensuremath{\varepsilon}}
\newcommand{\la}{\langle}
\newcommand{\ra}{\rangle}

\newcommand{\lemlab}[1]{\label{lemma:#1}}

\newcommand{\proplab}[1]{\label{prop:#1}}

\newcommand{\theolab}[1]{\label{theo:#1}}

\newcommand{\seclab}[1]{\label{sec:#1}}
\newcommand{\deflab}[1]{\label{def:#1}}

\def\lemref#1{Lemma~\ref{lemma:#1}}

\def\propref#1{Proposition~\ref{prop:#1}}

\def\theoref#1{Theorem~\ref{theo:#1}}

\def\secref#1{Section~\ref{sec:#1}}

\def\defref#1{Definition~\ref{def:#1}}

\newcommand{\U}{\mathcal{U}}
\DeclareMathOperator{\lca}{lca}

\title{Better Algorithms for Unfair Metrical Task Systems and
Applications\thanks{This work was partly supported by  United States Israel
Bi-national Science Foundation Grant 96-00247/1. Preliminary version appeared
in 32nd Annual ACM Symposium on Theory of Computing, 2000.
\copyright 2003 Society for Industrial and Applied Mathematics.}}

\author{
Amos Fiat\thanks{School of Computer Science, Tel-Aviv University, Tel-Aviv,
Israel (\texttt{fiat@tau.ac.il}).} \and Manor Mendel\thanks{School of
Computer Science, Tel-Aviv University, Tel-Aviv, Israel
(\texttt{mendelma@tau.ac.il}).}}
\begin{document}

\date{}

\maketitle

\begin{abstract}
Unfair metrical task systems are a generalization of online metrical task
systems. In this paper we introduce new techniques to combine algorithms for
unfair metrical task systems and apply these techniques to obtain
improved randomized online algorithms for metrical task systems
on arbitrary metric spaces.
\end{abstract}

\section{Introduction}

Metrical task systems, introduced by Borodin, Linial, and Saks \cite{BLS92},
can be described as follows: A server in some internal state receives
\emph{tasks} that have a service cost associated with each of the internal
states. The server may switch states, paying a cost given by a metric space
defined on the state space, and then pays the service cost associated with
the new state.

Metrical task systems have been the subject of a great deal of
study. A large part of the research into online algorithms can be
viewed as a study of some particular metrical task system. In
modelling some of these problems as metrical task systems, the set
of permissible tasks is constrained to fit the particulars of the
problem. In this paper we consider the original definition of
metrical task systems where the set of tasks can be arbitrary.

A deterministic algorithm for any $n$-state metrical task system with a
competitive ratio of $2n-1$ is given in~\cite{BLS92},
along with a matching lower bound for any metric space.

The \emph{randomized} competitive ratio of the MTS problem is not as well
understood. For the uniform metric space, where all distances are equal, the
randomized competitive ratio is known to within a constant factor, and is
$\Theta(\log n)$ \cite{BLS92,IS98}. In fact, it has been conjectured that
the randomized competitive ratio for MTS is $\Theta(\log n)$ in any
$n$-point metric space. Previously, the best upper bound on the competitive
ratio for arbitrary $n$-point metric space was $O(\log^5 n \log \log n)$ due Bartal,
Blum, Burch and Tomkins~\cite{BBBT97} and Bartal~\cite{Bar98}. The best lower
for any $n$-point metric space is $\Omega (\log n/ \log \log n)$ due to Bartal,
Bollob\'as and Mendel \cite{BBM01} and Bartal, Linial, Mendel and Naor \cite{BLMN03},
improving previous lower bounds of
Karloff, Rabani and Ravid \cite{KRR94}, and Blum, Karloff, Rabani, and Saks
\cite{BKRS00}.

As observed in \cite{KRR94,BKRS00,Bar96}, the randomized
competitive ratio of the  MTS is conceptually easier to analyze on
``decomposable spaces": spaces that have a partition to subspaces
with small diameter compared to that of the entire space.
Bartal~\cite{Bar96} introduced a class of decomposable spaces
called \emph{hierarchically well-separated trees} (HST).
Informally, a $k$-HST is a metric space having a partition into
subspaces such that: (i) the distances between the subspaces are
all equal; (ii) the diameter of each subspace is at most $1/k$
times the diameter of the whole space; and (iii) each subspace is
recursively a $k$-HST.

Following \cite{Bar96,BBBT97}, we obtain an improved algorithm for
HSTs. In order to reduce the MTS problem on arbitrary metric space
to a MTS problem on a HST we use probabilistic embedding of metric
spaces into HSTs \cite{Bar96}. It is shown in \cite{Bar98} that
any $n$-point metric space has probabilistic embedding in $k$-HSTs
with \emph{distortion} $O(k \log n \log\log n)$. Thus, an MTS
problem on an arbitrary $n$-point metric space, can be reduced to
an MTS problem on a $k$-HST with overhead of $O(k \log n \log \log
n)$ \cite{Bar96}.

Our algorithm for HSTs follows the general framework given
in~\cite{BKRS00} and explicitly formulated in~\cite{Sei99,BBBT97},
where the recursive structure of the HST is modelled by defining
an {\sl unfair metrical task system} problem \cite{Sei99,BBBT97}
on a uniform metric space. In an unfair MTS problem, associated
with every point $v_i$ of the metric space is a cost ratio $r_i$.
We charge the online algorithm a cost of $r_i c_i$ for dealing
with the task $(c_1,\ldots,c_i,\ldots,c_n)$ in state $v_i$.
 which
multiplies the online costs for processing tasks in that point.
Offline costs remain as before. The cost ratio $r_i$ roughly
corresponds to the competitive ratio of the online algorithm in a
subspace of the HST. For UMTSs on uniform metric spaces, tight
upper bounds are only known for two point spaces
\cite{BKRS00,Sei99,BBBT97} and for $n$ point spaces with equal
cost ratios \cite{BBBT97}. A tight lower bound is known for any
number of points and any cost ratios~\cite{BBM01}.

In this paper we introduce a general notation and technique for combining
algorithms for unfair metrical task systems on hierarchically decomposable
metric spaces. This technique is an improvement on the previous methods
\cite{BKRS00,Sei99,BBBT97}.
Using this
technique, we obtain randomized algorithms for unfair metrical task systems
on the uniform metric space that are better than the algorithm of
\cite{BBBT97}. Using the algorithm for unfair metrical task systems on
uniform metric space and the new method for combining algorithms, we obtain
$O(\log n \log \log n )$ competitive algorithms for MTS on HST spaces, which
implies $O((\log n \log\log n)^2)$-competitive randomized algorithm for
metrical task systems on any metric space.


We also study the \emph{weighted caching problem}. Weighted
caching is the paging problem when there are different costs to
fetch different pages. Deterministically, a competitive ratio of
$k$ is achievable~\cite{CKPV90,You91A}, with a matching lower
bound following from the $k$-server bound \cite{MMS90}. No
randomized algorithm is known to have a competitive ratio better
than the deterministic competitive ratio for general metric
spaces. However, in some special cases progress has been made.
Irani [personal communication] has shown an $O(\log k)$
competitive algorithm when page fetch costs are one of two
possible values. Blum, Furst, and Tomkins \cite{BFT96} have given
an $O(\log^2 k)$ competitive algorithm for arbitrary page costs,
when the total number of pages is $k+1$, they also present a lower
bound of $\Omega(\log k)$ for any page costs. As the weighted
caching problem with cache size $k$ on $k+1$ pages is a special
case of MTS on star-like metric spaces, we are able to obtain an
$O(\log k)$ competitive algorithm for this case,
improving~\cite{BFT96}. This is tight up to a constant factor.

\paragraph{Outline of the paper}
In \secref{prelim} the MTS problem is formally defined, along with
several technical conditions that later allow us to combine
algorithms for subspaces together. In \secref{mts-ub:combining} we
deal with the main technical contribution of our paper. We
introduce a novel technique to combine algorithms for subspace
into an algorithm for the entire space.
Section~\ref{section:building} is devoted for introducing algorithms for UMTSs
on uniform spaces.
In Section~\ref{sec:app} we give the applications mentioned
above by combining the algorithms of Section~\ref{section:building}.

\section{Preliminaries}
\label{sec:prelim}

\emph{Unfair metrical task systems} (UMTSs) \cite{Sei99,BBBT97} are a
generalization of metrical task systems \cite{BLS92}. A UMTS
$U=(M;(r_u)_{u\in M};s)$ consists of a metric space $M$ with a
distance metric $\dist_M$, a sequence of \emph{cost ratios} $r_u\in \Re^+$
for $u\in M$, and a \emph{distance ratio} $s\in \Re^+$.

Given a UMTS $U$, the associated online problem is defined as
follows. An online algorithm $A$ occupies some state $u \in M$.
When a task arrives the algorithm may change state to $v$. A task
is a tuple $(c_x)_{x\in M}$ of non-negative real numbers, and the
cost for algorithm $A$ associated with servicing the task is
$s\cdot \dist_M(u,v) + r_v c_v$. The cost for $A$ associated with
servicing a sequence of tasks $\sigma$ is the sum of costs for
servicing the individual tasks of the sequence consecutively. We
denote this sum by $\cost_A(\sigma)$. An online algorithm makes
its decisions based only upon tasks seen so far.

An off-line player is defined that services the same sequence of
tasks over $U$. The cost of an off-line player,
if it were to do exactly as above, would be $\dist_M(u,v) + c_v$. Thus, the
concept of \emph{unfairness}, the costs for doing the same thing are
different.

Given a sequence of tasks $\sigma$ we define the \emph{work function}
\cite{CL91} at $v$, $w_{\sigma,U}(v)$, to be the minimal cost, for any
off-line player, to start at the initial state in $U$, deal with all tasks
in $\sigma$, and end up in state $v$. We omit the use of the subscript $U$
if it is clear from the context. Note that for all $u, v\in M$,
$w_\sigma(u)-w_\sigma(v) \leq \dist_M(u,v)$. If $w_\sigma(u)=w_\sigma(v) +
\dist_M(u,v)$, $u$ is said to be \emph{supported} by $v$. We say that $u\in
M$ is \emph{supported} if there exists some $v\in M$ such that $u$ is
supported by $v$.

We define $\cost_{\text{OPT}}(\sigma)$ to be $\min_v w_\sigma(v)$.
This is simply the minimal cost, for any off-line player, to start
at the initial state and process $\sigma$. As the differences
between the work function values on different states is bounded by
a constant (the diameter of the metric space) independent of the
task sequence, it is possible to use a convex combination of the
work function values instead of the minimal one. We say that
$\alpha=(\alpha(u))_{u\in M}$ is a \emph{weight vector} when
$\{\alpha(u)| u\in M \}$ are  non-negative real numbers satisfying
$\sum_{u\in M} \alpha(u)=1$. We define the $\alpha$-optimal-cost
of a sequence of tasks $\sigma$ to be $\cost_{\aopt}(\sigma)=$
$\la\alpha, w_{\sigma}\ra$ $= \sum_{u\in M} \alpha(u)
w_\sigma(u)$. As observed above, $\cost_{\aopt}(\sigma) \leq
\cost_{\text{OPT}}(\sigma)+ \diam (M)$, where
$\diam(M)=\max_{u,v\in M} \dist_M(u,v)$ is the diameter of $M$.

A randomized online algorithm $A$ for a UMTS is a probability
distribution over deterministic online algorithms. The expected
cost of a randomized algorithm $A$ on a sequence $\sigma$ is
denoted by $E[\cost_A(\sigma)]$.
\begin{definition} \cite{SleTar85a,KMRS88,BBKTW94} \label{def:competitive}
A randomized online algorithm $A$ is called $r$ competitive
against an oblivious adversary if  there exists some $c$ such that
for all task sequences $\sigma$, $E[\cost_A(\sigma)] \leq r
\,\cost_{OPT}(\sigma) + c$.
\end{definition}

\begin{observation} \label{obs:w/o-dr}
We can limit the discussion on the competitive ratio of UMTSs to distance ratio equals one since a UMTS
$U=(M;(r_u)_{u\in M};s)$ has a competitive ratio of $r$ if and only if
$U'=(M; (s^{-1}r_u)_{u\in M};1)$ has competitive ratio of $r s^{-1}$.
Moreover an $rs^{-1}$ competitive algorithm for $U'$ is $r$ competitive
algorithm for $U$, since in both $U'$ and $U$ the offline costs are the same
but the online costs in $U$ are multiplied by a factor of $s$ compared to
the costs in $U'$. When $s=1$, we drop it from the notation.
\end{observation}

Given a randomized online algorithm $A$ for a UMTS $U$ with state space $M$
and a sequence of tasks $\sigma$, we define $p_{\sigma,A}$ to be the vector
of probabilities $(p_{\sigma,A}(u))_{u\in M}$ where $p_{\sigma,A}(u)$ is the
probability that $A$ is in state $u$ after serving the request sequence
$\sigma$. We drop the subscript $A$ if the algorithm is clear from the
context.

Let $x\circ y$ denote the concatenation of sequences $x$ and $y$. Let $U$ be
a UMTS over the metric space $M$ with distance ratio $s$. Given two
successive probability distributions on the states of $U$, $p_{\sigma}$ and
$p_{\sigma\circ e}$, where $e$ is the next task, we define the set of
transfer matrices from $p_{\sigma}$ to $p_{\sigma \circ e}$, denoted
$T(p_{\sigma},p_{\sigma\circ e})$, as the set of all matrices $T=(t_{u
v})_{u,v \in M}$ with non negative real entries, where
\begin{align*}
\sum_{v\in M} t_{u v} &= p_{\sigma}(u), \ u \in M;
&\sum_{u\in M} t_{u v} &= p_{\sigma\circ e}(v), \ v \in M.
\end{align*}
We define the \emph{unweighted moving cost} from $p_{\sigma}$ to
$p_{\sigma\circ e}$:
\begin{equation*}
\mcost_M(p_{\sigma},p_{\sigma\circ e}) =  \negthickspace
 \min_{\begin{smallmatrix} (t_{u v})\in \\ T(p_{\sigma},p_{\sigma\circ e})\end{smallmatrix}}
  \sum_{u,v} t_{u v} \dist_M(u,v),
\end{equation*}
the \emph{moving cost} 
is defined as $\mcost_U(p_\sigma,p_{\sigma\circ e})=$ $s\cdot
\mcost_M(p_\sigma,$ $ p_{\sigma\circ e})$, and the \emph{local cost} on a
task  $e=(c_u)_{u\in M}$ is defined as $\sum_{u \in M} p_{\sigma\circ e}(u)
c_u r_u$. Due to linearity of expectation, $E[\cost_A(\sigma\circ e)] -
E[\cost_A(\sigma)]$ is equal to the sum of the moving cost from $p_{\sigma}$ to
$p_{\sigma\circ e}$ and the local cost on $e$. Hence we can view $A$ as a
deterministic algorithm that maintains the probability mass on the states
whose cost on task $e$ given after sequence $\sigma$ is
\begin{equation} \label{eq:oncost}
\cost_A(\sigma\circ e) - \cost_A(\sigma)=  \mcost_U(p_\sigma,p_{\sigma \circ
e}) + \sum_{u \in M} p_{\sigma \circ e}(u) c_u r_u.
\end{equation}
In the sequel we will use the terminology of changing probabilities, with the
understanding that we are referring to a deterministic algorithm charged
according to~\eqref{eq:oncost}.

\medskip

We next develop some technical conditions that make it easier to
combine algorithms for UMTSs. \emph{Elementary tasks} are tasks
with only one non-zero entry, we use the notation $(v,\delta)$,
$\delta\geq 0$, for an elementary task of cost $\delta$ at state
$v$. Tasks $(v,0)$ can simply be ignored by the algorithm.

\begin{definition}[\cite{BBBT97}]
A \emph{reasonable} algorithm is an online algorithm that never assigns a
positive probability to a supported state.
\end{definition}

\begin{definition}[\cite{BBBT97}]
A \emph{reasonable task sequence} for algorithm $A$, is a sequence of tasks
that obeys the following:
\begin{enumerate}
\item All tasks are elementary.
\item For all $\sigma$, the next task $(v,\delta)$ must obey that for all
$\delta'$, if $\delta>\delta'\geq 0$ then $p_{\sigma\circ (v,\delta')}(v)>0$.
\end{enumerate}
\end{definition}

\noindent It follows that a reasonable task sequence for $A$ never includes
tasks $(v,\delta)$, $\delta>0$, if the current probability of $A$ on $v$ is
zero.

The following lemma is from \cite{BBBT97}. For the sake of completeness, we include a sketch of
a proof here.
\begin{lemma}
\label{th:reasonableadv} Given a randomized online algorithm $A_0$ that
obtains a competitive ratio of $r$ when the task sequences are limited to
being reasonable task sequences for $A_0$, then, for all $\e>0$, there also
exists a randomized algorithm $A_3$ that obtains a competitive ratio of
$r+\e$ on all possible sequences.
\end{lemma}
\begin{proof}[sketch]
The proof proceeds in three stages. In the first stage, we convert an algorithm $A_0$
for reasonable task sequences to a \emph{lazy} algorithm $A_1$ (an algorithm
that dose not move the server when receiving a task with zero cost) for
reasonable task sequences. In the second stage, we convert an algorithm
$A_1$ to an algorithm $A_2$ for elementary task sequences, and then, in the
third stage, we convert $A_2$ to an algorithm $A_3$ for general task
sequences.

The first stage is well known.

The second stage. Given an elementary task sequence, every elementary task
$e=(v,x)$ is converted to a task $(v,y)$ such that $y = \sup \{ z | z<x$ and
the probability induced by $A_1$ on $v$ is greater than $0 \}$. The resulting
task sequence is reasonable and is fed to $A_1$. $A_2$ imitates the movements
of $A_1$.

The third stage. Let $\sigma$ be an arbitrary task sequence. First, we
convert $\sigma$ into an elementary task sequence $\hat{\sigma}$, each task
$\tau=( \delta_1,\ldots,\delta_n)$ in $\sigma$ is converted to a sequence of
tasks $\hat{\sigma}_\tau$ as follows: Let $\e'>0$ be small constant to be
determined later, and assume for simplicity that $\delta_i\geq \delta_{i+1}$.
Then \( \hat{\sigma}_\tau= \varsigma_1\circ \varsigma_2 \circ \cdots
\varsigma_N, \) where  $N= \lfloor \delta_1 / \e' \rfloor$ and
 \( \varsigma_j= (v_1,\e') \circ (v_2,\e') \circ \cdots \circ
 (v_{k_j},\e'), \)
where $k_j= \max \{ i | \delta_i \geq j\cdot \e' \}$. Note that the optimal
offline cost $\hat{\sigma}$ is at most the optimal offline cost on $\sigma$,
since any servicing for $\sigma$, when applied to $\hat{\sigma}$ would have
a cost no bigger than the original cost. Consider an $r$-competitive online
algorithm $A_2$ for elementary tasks operating on $\hat{\sigma}$, and
construct an online algorithm $A_3$ for $\sigma$. $B$ maintains the invariant
that the state of $A_3$ after processing some task $\tau$ is the same state as
$A_2$ after processing the sequence $\hat{\sigma}_\tau$. Consider the behavior
of $A_2$ on $\hat{\sigma}_\tau$. It begins in some state $v_{i_0}$, passes
through some set $S$ of states and ends up in some state $v_{i_2}$. Consider
the original task $\tau=(\delta_1,\ldots,\delta_n)$. Let $v_{i_1}$ be the
state in $S$ with the lowest cost in $\tau$. Algorithm $A_3$ begins in state
$v_{i_0}$, immediately moves to $v_{i_1}$, serves $\tau$ in $v_{i_1}$ and
then moves to $v_{i_2}$.

Informally, on each task $A_2$ pays either a local cost of $\e'$ or moving
cost of at least $\e'$ and therefore these costs are larger than the local
cost of $A_3$. $A_3$ also has a moving cost at least as $A_2$. By a careful
combination of these two we can conclude that the cost of $B$ on $\sigma$ is
at most $(1+\e)$ times the cost of $A_2$ on $\hat{\sigma}$.
\end{proof}

\emph{Hereafter, we assume only reasonable task sequences}. This is without
lost of generality due to Lemma~\ref{th:reasonableadv}.

\begin{observation}
\label{obser:ob2} When a reasonable algorithm $A$ is applied to a
reasonable task sequence $\sigma=\tau_1\tau_2\cdots \tau_m$, any elementary
task $\tau=(v,\delta)$ causes the work-function at $v$, $w(v)$, to increase by $\delta$. This follows
because $v$ would not have been supported following any alternative request
$(v,\delta')$, $\delta'<\delta$. See~\cite[Lemma~1]{BBBT97} for a rigorous treatment.
This also implies that for any state  $v$, $w_\sigma(v)= \sum_{j=1}^m \tau_j(v)$.
\end{observation}

\begin{definition} \label{definition:resonable}
An online algorithm $A$ is said to be \emph{sensible} and
$r$-competitive on the UMTS $U=(M;(r_u)_{u\in M};s)$  if it obeys
the following:
\begin{enumerate}
\item $A$ is reasonable.
\item $A$ is a \emph{stable algorithm} \cite{CL91}, {\ie}, the probabilities that $A$ assigns to
the different states are purely a function of the work function.
\item \label{definition:resonable:potential}
Associated with $A$ are a weight vector $\alpha_A$ and a potential function
$\Phi_A$ such that
\begin{itemize}
\item $\Phi_A:\Re^b\mapsto \Re^+$, is purely a function of the work-function, bounded, non-negative, and
continuous.
\item For all task sequences $\sigma$ and all tasks $e$,
\begin{equation}
\cost_{A}(\sigma\circ e)-\cost_A(\sigma) + \Phi_A(w_{\sigma \circ e}) -
\Phi_A({w}_{\sigma})
 \leq r \cdot \la\alpha_A, w_{\sigma \circ e} - w_{\sigma}\ra.
 \label{eq:reasonable:potential}
\end{equation}
\end{itemize}
\end{enumerate}
\end{definition}

\begin{observation}
An online algorithm that is sensible and $r$-competitive (against reasonable
task sequences) according to
Def.~\ref{definition:resonable} is also $r$-competitive according to
Def.~\ref{def:competitive}. This is so since summing up the two sides in
Inequality~\eqref{eq:reasonable:potential} over the individual tasks in the
task sequence, we get a telescopic sum such that \( \cost_{A}(\sigma) +
\Phi_A(w_{\sigma }) - \Phi_A({w}_{\e})
 \leq r \cdot \la\alpha_A, w_{\sigma } - w_{\e} \ra ,\)
where $w_{\e}$ is the initial work function. We conclude that \(
\cost_{A}(\sigma) \leq r \cdot \cost_{\text{OPT}}(\sigma)+ r\Delta(M)
+\sup_w \Phi(w) .\)
\end{observation}

When combining sensible algorithms we would like the resulting algorithm to
be also sensible. The problematic invariant to maintain is reasonableness.
In order to maintain reasonableness there is a need for a stronger concept,
which we call \emph{constrained algorithms}.

\begin{definition} \deflab{betaetacons}
A sensible  $r$-competitive algorithm $A$ for the UMTS
$U=(M;(r_u)_{u\in M};s)$ with associated potential function $\Phi$
is called $(\beta,\eta)$-constrained, $0\leq \beta\leq 1$, $0\leq
\eta$, if the following hold:
\begin{enumerate}
\item For all $u,v\in M$: if $w(u)-w(v)\geq \beta \dist_M(u,v)$ then the
probability that $A$ assigns to $u$ is zero ($p_{w,A}(u)=0$).
\item $\|\Phi\|_\infty \leq \eta\, \diam(M)r$, where $\|\Phi\|_\infty =\sup_{w}
\Phi(w)$.
 \end{enumerate}
\end{definition}

\begin{observation} \label{obs:constrained}
\begin{enumerate}
\item  For a $(\beta,\eta)$-constrained algorithm competing against a reasonable
task sequence,
 \(  \forall u, v\in M,\;  |w(u)-w(v)|\leq \beta\,\dist_M(u,v) .\)
The argument here is similar to the one given in Observation~\ref{obser:ob2}.

\item A sensible $r$-competitive algorithm for a metric space of diameter $\Delta$
is by definition a $(1,|\Phi_A|/(r\Delta))$-constrained.

\item A $(\beta,\eta)$-constrained algorithm is trivially
$(\beta',\eta')$-constrained for all $\beta\leq \beta' \leq 1$ and $\eta \leq
\eta'$.
\end{enumerate}
\end{observation}

\section{A Combining Theorem for Unfair Metrical Task Systems}
\seclab{mts-ub:combining}

Consider a metric space $M$ having a partition to sub-spaces $M_1,\ldots, M_b$,
with ``large" distances between sub-spaces compared to the diameters of the
sub-spaces. A metrical task system on $M$ induces metrical task systems on
$M_i$, $i\in\{1,\ldots,b\}$. Assume that for every $i$, we have a $\hat{r}_i$-competitive
algorithm $A_i$ for the induced MTS on $M_i$. Our goal is to combine the
$A_i$ algorithms so as to obtain an algorithm for the original MTS defined
on $M$. To do so we make use of a ``combining algorithm" $\hat{A}$.
$\hat{A}$ has the role of determining which of the $M_i$ sub-spaces contains
the server. Since the ``local cost" of $\hat{A}$ on sub-space $M_i$ is
$\hat{r}_i$ times the optimal cost on subspace $M_i$, it is natural that
$\hat{A}$ should be an algorithm for the UMTS $\hat{U}=(\hat{M};
(\hat{r}_1,\ldots,\hat{r}_b)),$ where $\hat{M}=\{z_1,\ldots,z_b\}$ is a space
with points corresponding to the sub-spaces and distances that are roughly
the distances between the corresponding sub-spaces. Tasks for $M$
are translated to tasks for the $M_i$ induced metrical task systems
simply by restriction. It remains to define how one translates tasks for $M$
to tasks for $\hat{U}$.

Previous papers \cite{BKRS00,Sei99,BBBT97} use the cost of the
optimal algorithm for the task in the sub-space $M_i$ as the cost
for $z_i$ in the task for $\hat{U}$. This way the local cost for
$\hat{A}$ is $\hat{r}_i$ times the cost for the optimum, however,
\emph{this is true only in the amortized sense}. In order to bound
the fluctuation around the amortized cost, those papers have to
assume that the diameters of the sub-space are very small compared
to the distances between $M_i$ sub-spaces. We take a different
approach: the cost for a point $z_i\in \hat{U}$ is (an upper bound
for) \emph{the cost of $A_i$ on the corresponding task, divided by
$\hat{r}_i$}. In this way the amortization problem disappears, and
we are able to combine sub-spaces with a relatively large
diameter. A formal description of the construction is given below.

\begin{theorem}\label{th:Aisrhat}
Let $U$ be a UMTS $U=(M;(r_u)_{u\in M};s)$, where $M$ is a metric
space on $n$ points. Consider a partition of the points of $M$,
$P=(M_1, M_2, \ldots, M_b)$. $U_j=(M_j;(r_u)_{u \in M_j};s)$ is
the UMTS induced by $U$ on the subspace $M_j$. Let $\hat{M}$ be a
metric space defined over the set of points $\{z_1, z_2, \ldots,
z_b\}$ with a distance metric
$\dist_{\hat{M}}(z_i,z_j)\geq\max\{\dist_M(u,v) : u\in M_i, v \in
M_j\}$. Assume that
\begin{itemize}
\item
For all $j$, there  is a $(\beta_j,\eta_j)$-constrained
$\hat{r}_j$-competitive algorithm $A_j$ for the UMTS $U_j$.
\item There is a
$(\hat{\beta},\hat{\eta})$-constrained $r$-competitive algorithm $\hat{A}$
for the UMTS $\hat{U}=(\hat{M};(\hat{r}_1, \ldots, \hat{r}_b);s)$.
\end{itemize}

Define
\begin{equation}\label{eqbeta}
\beta = \max \Bigl\{\max_i\beta_i,\;  \max_{i\neq j}
\frac{\hat{\beta}\dist_{\hat{M}}(z_i,z_j) + \beta_j\,\diam(M_j) +
\beta_i\,\diam(M_i) + \eta_i \diam(M_i)}{\min_{p\in M_i, q\in
M_j}\dist_M(p,q)}\Bigr\},
\end{equation}
and
\begin{equation}\label{eqeta}
 \eta = \hat{\eta}\frac{\diam(\hat{M})}{\diam(M)}+\max_i
\, \eta_i \frac{ \diam(M_i)}{\diam(M)}.
\end{equation}

If $\beta \leq 1$, then there exists a $(\beta,\eta)$-constrained
and $r$-compe\-ti\-tive algorithm, $A$, for the UMTS $U$.
\end{theorem}

In our applications of Theorem~\ref{th:Aisrhat}, the metric space
$M$ have a``nice" partition $P=(M_1,\ldots,M_b)$, parameterized
with $k\geq 1$: $d_M(u,v)=\diam(M)$ for all $i\neq j$ $u\in M_i$,
$v\in M_j$; and $\diam(M_i)\leq \diam(M)/k$. In this case the
statement of Theorem~\ref{th:Aisrhat} can be simplified as
follows.

\begin{corollary} \label{cor:Aisrhat}
Under the assumptions of Theorem~\ref{th:Aisrhat}, and assuming
the partition is ``nice" (with parameter $k$), in the above sense.
Define
\begin{equation} \label{eqbeta2}
  \beta=\max\{ \max_i \beta_i, \hat{\beta} +\frac{\max_{i\neq j} (\beta_i+\beta_j+ \eta_i)}{k} \},
\end{equation}
and
\begin{equation} \label{eqeta2}
\eta = \hat{\eta}+ \frac{\max_i \eta_i}{k} .
\end{equation}
If $\beta \leq 1$, then there exists a $(\beta,\eta)$-constrained
and $r$-compe\-ti\-tive algorithm, $A$, for the UMTS $U$.
\end{corollary}

\medskip
In Section~\ref{sec:construct} we define the combined algorithm
$A$ declared in Theorem~\ref{th:Aisrhat}.
Section~\ref{section:ProofAisrhat} contains the proof of
Theorem~\ref{th:Aisrhat}. We end the discussion on the combining
technique with Section~\ref{sec:constrained} in which we show how
to obtain constrained algorithms needed in the assumptions of
Theorem~\ref{th:Aisrhat}.

\subsection{The Construction of the Combined Algorithm} \label{sec:construct}

Denote by $\Phi_j$ and  $\alpha_j$  the associated
potential function and weight vector  of algorithm $A_j$, respectively.
Similarly, denote by $\hat{\Phi}$ and $\hat{\alpha}$  the associated
potential function and weight vector of algorithm $\hat{A}$, respectively.

Given a
sequence of elementary tasks $\sigma = (v_1,\delta_1)\circ (v_2,
\delta_2)\circ \dots \circ (v_{|\sigma|},\delta_{|\sigma|})$, $v_i\in M$, we
define the sequences
\begin{equation*}
\sigma |_{M_\ell}=(u^\ell_1,\delta^\ell_1) \circ (u^\ell_2,\delta^\ell_2)
\circ \dots \circ (u^\ell_{|\sigma|}, \delta^\ell_{|\sigma|}), \text{ where }
\end{equation*}
\begin{itemize}
\item $u^\ell_j = v_j$ and $\delta_j^\ell = \delta_j$, if $v_j\in M_\ell$.
\item $u^\ell_j$ is an arbitrary point in $M_\ell$ and $\delta_j^\ell=0$, if $v_j\notin
M_\ell$.
\end{itemize}
Informally, $\sigma|_{M_\ell}$ is the restriction of $\sigma$ to subspace
$M_\ell$.


\noindent For $u\in M$, define $s(u)=i$ if and only if $u\in M_i$.
 We define the sequence
 \[\chi(\sigma) = (z_{s(v_1)},\hat{\delta}_1) \circ (z_{s(v_2)},
\hat{\delta}_2) \circ \dots \circ
(z_{s(v_{|\sigma|})},\hat{\delta}_{|\sigma|}),\] inductively. Let
$e=(v,\delta)$, $s(v)=\ell$, then $\chi(\sigma\circ e)=\chi(\sigma)\circ
(z_\ell,\hat{\delta})$ where
\begin{equation} \label{eq:hatdelta}
\hat{\delta} =
 \bigl( \la {\alpha}_{\ell},{w_{(\sigma\circ e)|_{M_\ell},U_\ell}}\ra
  - \Phi_{\ell}(w_{(\sigma\circ e)|_{M_\ell},U_\ell})/\hat{r}_\ell \bigr)
  -
 \bigl( \la {\alpha}_{\ell},{w_{\sigma|_{M_\ell},U_\ell}}\ra
  - \Phi_{\ell}(w_{\sigma|_{M_\ell},U_\ell})/\hat{r}_\ell \bigr).
\end{equation}

Note that  $\hat{\delta}$ is an upper bound on the cost of
$A_\ell$ for the task $(v,\delta)$, divided by $\hat{r}_\ell$.
This fact follows from \eqref{eq:reasonable:potential} since
$A_\ell$ is sensible, and $\sigma | _{M_\ell}$ is a reasonable
task sequence for $A_\ell$ (see \lemref{resadv}). It also implies
that $\hat{\delta}\geq 0$, which is a necessary requirement for
$(z_\ell ,\hat{\delta})$ to be a well defined task.

\begin{combined}
The algorithm works as follows:
\begin{enumerate}
\item It simulates algorithm $A_\ell$ on
the task sequence $\sigma|_{M_\ell}$, for $1\leq \ell \leq b$.
\item It also simulates algorithm $\hat{A}$ on the task sequence $\chi(\sigma)$.
\item The probability assigned to a
point $v\in M_\ell$ is the product of the probability assigned by $A_\ell$ to
$v$ and the probability assigned by $\hat{A}$ to $z_\ell$. (\ie,
$p_{\sigma,A}(v)=p_{\sigma|_{M_\ell},A_{\ell}}(v)\cdot
p_{\chi(\sigma),\hat{A}}(z_\ell)$.)
\end{enumerate}
\end{combined}

\medskip

We remark that the simulations above can be performed in an online fashion.

\subsection{Proof of Theorem \ref{th:Aisrhat}}\label{section:ProofAisrhat}

 To simplify notation we use the following shorthand notation. Given a task
sequence $\sigma$ and a task $e$. With respect to $\sigma$, we
define
\begin{align*}
w&=w_{\sigma,U}; & w^e&=w_{\sigma\circ e,U};
\\
w_k&=w_{\sigma|_{M_k},U_k}, \  1\leq k \leq b; & w_{k}^e&=w_{(\sigma\circ
e)|_{M_k},U_k}, \  1\leq k \leq b;
\\
\hat{w}&=w_{\chi(\sigma),\hat{U}}; & \hat{w}^e&=w_{\chi(\sigma\circ
e),\hat{U}}.
\end{align*}

Define $p$, $p_k$, and $\hat{p}$ to be the probability
distributions on the states of $U$, $U_k$ and $\hat{U}$ as induced
by algorithms $A$, $A_k$ and $\hat{A}$ on the sequences $\sigma$,
$\sigma|_{M_k}$, and $\chi(\sigma)$, $1\leq k \leq b$,
respectively. Likewise, we define $p^e$, $p_k^e$ and $\hat{p}^e$
where the sequences are $\sigma\circ e$, $\sigma\circ e|_{M_k}$,
and $\chi(\sigma\circ e)$.



\begin{lemma} \lemlab{resadv}
If the task sequence $\sigma$  given to algorithm $A$ on $U$ is reasonable, then the
simulated task sequences $\sigma|_{M_i}$ for algorithms $A_i$ on $U_i$ and the simulated
task sequence $\chi(\sigma)$ for algorithm $\hat{A}$ on $\hat{U}$ are also reasonable.
\end{lemma}
\begin{proof}
We first prove that $\sigma'|_{M_\ell}$ is reasonable for $A_\ell$
by induction on $|\sigma'|$. Say $\sigma'=\sigma \circ e$,
$e=(v,\delta)$, and $v \in M_\ell$. Since $\sigma'$ is reasonable
for $A$, would the task $e$ have been replaced with the task
$e'=(v,\delta')$, and $\delta' \in [0,\delta)$, then by the
reasonableness of $\sigma'$,  $p^{e'}(v)>0$, but since
$p^{e'}(v)=p^{e'}_\ell(v) \hat{p}^{e'}(z_\ell)$ it follows that
$p^{e'}_\ell(v)>0$. This implies $\sigma'|_{M_\ell}$ is reasonable
for $A_\ell$.

We next prove that $\chi(\sigma')$ is a reasonable task sequence
for $\hat{A}$, by induction on $|\sigma'|$. Let
$\sigma'=\sigma\circ e$, $e=(v,\delta)$, $v\in M_\ell$. Denote by
$\hat{e}=(z_{\ell},\hat{\delta})$ the last task in $\chi(\sigma)$.
Consider a hypothetical task $(v,x)$ in $U$, for $0 \leq x \leq
\delta$. Denote by $(z_\ell,f(x))$ the corresponding task for
$\hat{U}$, where $f(x)$ is determined according to
\eqref{eq:hatdelta}. $f$ is continuous (since $\Phi_{\ell}$ is
continuous), $f(0)=0$, and $f(\delta)=\hat{\delta}$. Therefore for
any $0 \leq \hat{\delta}' < \hat{\delta}$ there exists $0 \leq
\delta' <\delta$ such that $f(\delta')=\hat{\delta}'$ and since
$0<p^{(v,\delta')}=p^{(v,\delta')}_\ell(v) \cdot
\hat{p}^{(v,\delta')}(z_\ell)$ we conclude that
$0<\hat{p}^{(v,\delta')}(z_\ell)$ (the probability induced by
$\hat{A}$ on $z_\ell$ after the task $(z_\ell,\hat{\delta}')$).
This implies that $\chi(\sigma)$ is a reasonable task sequence for
$\hat{A}$.
\end{proof}

\begin{lemma} \lemlab{hatw}
For all $\sigma$ and for all $\ell$, $\hat{w}(z_{\ell}) = \la
{\alpha}_{\ell},{w_\ell}\ra  - \Phi_{\ell}({w_\ell})/\hat{r}_\ell.$
\end{lemma}
\begin{proof}
 It follows from \lemref{resadv} that the task sequence $\chi(\sigma)$ for
$\hat{A}$ is reasonable. As $\hat{A}$ is sensible it follows from
Observation~\ref{obser:ob2} that $\hat{w}(z_\ell)$ is exactly the
sum of costs in $\chi(\sigma)$ for $z_\ell$. By the definition of
$\chi(\sigma)$ in (see \eqref{eq:hatdelta}) it follows that this sum is
$\la {\alpha}_{\ell},{w_\ell}\ra  - \Phi_{\ell}({w_\ell})/\hat{r}_\ell$.
\end{proof}

\begin{lemma} \lemlab{betatagc}
Assume that $w(u)=w_\ell(u)$
for all $1\leq \ell\leq b$, $u\in M_\ell$.
Then any state $u\in U$ for which there exists a state $v$ such that
$w(u)-w(v) \geq \beta \dist_M(u,v)$, has $p(u)=0$.
\end{lemma}
\begin{proof}
Consider states $u$ and $v$ as above, \ie, $w(u)-w(v) \geq \beta
\dist_M(u,v)$. We now consider two cases:
\begin{enumerate}
\item $u,v\in M_i$. We want to show that $w_i(u)-w_i(v)\geq
\beta_i \dist_{M_i}(u,v)$, as $A_i$ is $(\beta_i,\eta_i)$-constrained this
implies that $p_i(u)=0$, which implies that $p(u)=0$. From the conditions
above we get
\begin{equation*}
 w_i(u)-w_i(v) = w(u)-w(v)
 \geq \beta \dist_M(u,v)  \geq \beta_i \dist_{M_i}(u,v).
\end{equation*}

\item $u\in M_i$, $v\in M_j$, $i\neq j$. Our
goal now will be to show that $\hat{w}(z_i) - \hat{w}(z_j) \geq
\hat{\beta}\dist_{\hat{M}} (z_i, z_j)$, as this implies that $\hat{p}(z_i)=0$
which implies that $p(u)=0$.

A lower bound on $\hat{w}(z_i)$ is
\begin{eqnarray}
\hat{w}(z_i) &=&  \la \alpha_i,w_i \ra - \|\Phi_i\|_\infty /r_i \label{eq1}\\
&\geq& w_i(u)-\beta_i \diam(M_i) - |\Phi_i|/r_i \label{eq2}\\
 &=& w(u)-\beta_i \diam(M_i) - \eta_i\diam(M_i). \label{eq3}
\end{eqnarray}
To justify \eqref{eq1} one uses the definitions and \lemref{hatw}.
Inequality~\eqref{eq2} follows because a convex combination of
values is at least one of these values minus the maximal
difference. The maximal difference between work function values is
bounded by $\beta_i$ times the distance, see
Observation~\ref{obs:constrained}. Equation~\eqref{eq3} follows
from our assumption that the work functions are equal and from the
definition of $\eta_i$.

Similarly, to obtain an upper bound on $\hat{w}(z_j)$, we derive
\begin{equation}
\hat{w}(z_j)
 = \la \alpha_j, w_j \ra - \|\Phi_j\|_\infty /r_j \leq w(v) + \beta_j \diam(M_j). \label{eq4}
\end{equation}

It follows from \eqref{eq3} and \eqref{eq4} that,
\begin{multline*}
\hat{w}(z_i) - \hat{w}(z_j)
 \geq (w(u)-w(v)) -\beta_i\diam(M_i) - \beta_j\diam(M_j) - \eta_i\diam{M_i} \\
\geq \beta d_M(u,v)-\beta_i\diam(M_i) - \beta_j\diam(M_j) - \eta_i\diam{M_i}
 \geq \hat{\beta}\dist_{\hat{M}}(z_i,z_j).
 \end{multline*}
The last inequality follows from \eqref{eqbeta}.
\end{enumerate}
\end{proof}

\begin{lemma} \lemlab{welleqw}
 For any reasonable task sequence $\sigma$, subspace $M_\ell$, and $v\in M_\ell$ it holds that
$w_\ell(v)=w(v)$.
\end{lemma}
\begin{proof}
Assume the contrary. Let $\sigma'$ be the shortest reasonable
task sequence for which there exists $v\in M_\ell$ satisfying
$w_{\sigma'|_{M_\ell}}(v)\neq w_{\sigma'}(v)$. It is easy to observe that
$\sigma'=\sigma\circ e$ where $e=(v,\delta)$. As the sequence $(\sigma\circ e)
|_{M_\ell}$ is a reasonable task sequence (\lemref{resadv}) and
$A_\ell$ is reasonable, it follows that $w^e_\ell(v) = w_\ell(v) + \delta$.
Since $w_\ell(v)=w(v)$ and $w^e(v) \leq w(v)+\delta$ we deduce that
$w^e_{\ell}(v)> w^e(v)$.

Let $e_x = (v,x)$, define \( \delta' = \sup \{x:
w^{e_x}(v)=w_\ell^{e_x}(v)\}\). Obviously, $0\leq \delta' \leq \delta$.
Define $e'=(v,\delta')$. By continuity of the work function
$w^{e'}(v)=w_\ell^{e'}(v)$ and thus $\delta'<\delta$. The conditions above
imply that an elementary task in $v$ after $w^{e'}$ will not change the work
function, which means that $v$ is supported in $w^{e'}$. Hence, the
assumptions of \lemref{betatagc} are satisfied (here we use the assumption that
$\beta \leq 1$). By \lemref{betatagc} $p^{e'}(v)=0$ and since the sequence
$\sigma$ is reasonable for $A$ it follows that $\delta \leq \delta'$, a
contradiction.
\end{proof}

\begin{proposition} \label{lemma:samecompratio}
 For all $\sigma$, and all tasks $e=(v,\delta)$,
\begin{equation*}\cost_{A}(\sigma\circ e) - \cost_A(\sigma)
\leq \cost_{\hat{A}} (\chi(\sigma\circ e)) - \cost_{\hat{A}}(\chi(\sigma)).
\end{equation*}
\end{proposition}
\begin{proof}
Let us denote the subspace containing $v$ by $M_\ell$.  We split the cost of
$A$ into two main components, the moving cost $\mcost_U(p,p^e)$, and the
local cost $r_v p^e(v) \delta=r_v \hat{p}^e(z_\ell)p_\ell(v_i)\delta$ (see
Equation \eqref{eq:oncost}).

We give an upper bound on the moving cost of $A$ by considering a possibly
suboptimal algorithm that works as follows:
\begin{enumerate}
\item Move probabilities between the different $M_j$ subspaces.
\emph{I.e.}, change the probability $p(u)=\hat{p}(z_j) p_j(u)$ for $u\in M_j$
to an intermediate stage $\hat{p}^e(z_j)p_j(v)$. The moving cost for $A$ to
produce this intermediate probability is bounded by
$\mcost_{\hat{U}}(\hat{p},\hat{p}^e)$  as the distances in $\hat{M}$ are an
upper bound on the real distances for $A$ ($\dist_{\hat{M}}(z_i,z_j) \geq
\dist_M(u,v)$ for $u\in M_i$, $v\in M_j$). We call this cost the inter-space
cost for $A$.
\item Move probabilities within the $M_j$ subspaces. {\sl I.e.}, move from the
intermediate probability $\hat{p}^e(z_j)p_j(u)$, $u\in M_j$ to the
probability $p^e(u)=\hat{p}^e(z_j)p_j^e(u)$. As all algorithms $A_j$, $j\neq
\ell$, get a task of zero cost, $p_j^e = p_j$, $j \neq \ell$. The moving cost
for $A$ to produce $p^e(u)$, $u\in M_\ell$, from the intermediate stage , is
no more than $\hat{p}^e(z_\ell)\cdot \mcost_{U_\ell}(p_\ell, p^e_\ell)$. We
call this cost the intra-space cost for $A$.
\end{enumerate}

Taking the local cost for $A$ and the intra-space cost for $A$:
\begin{align}
r_u \hat{p}^e(z_\ell)& p_\ell(u)\delta +\hat{p}^e(z_\ell)\cdot
\mcost_{U_\ell}(p_\ell, p^e_\ell)\nonumber \\
&= \hat{p}^e(z_\ell) \left(cost_{A_\ell}(\sigma \circ e) -
\cost_{A_\ell}(\sigma) \right) \label{eq32}
\\ &\leq \hat{p}^e(z_\ell) \hat{r}_\ell \bigl( (\la \alpha_\ell, w^e_\ell \ra -
\Phi_\ell(w^e_\ell)/\hat{r}_\ell)   - (\la \alpha_\ell, w_\ell \ra -
\Phi_\ell(w_\ell)/\hat{r}_\ell) \bigr) \label{eq33} 
\end{align}

To obtain \eqref{eq32} we use the definition of online cost (see
\eqref{eq:oncost}). To obtain \eqref{eq33} we use the fact that $A_\ell$
is $\hat{r}_\ell$ competitive and sensible (see
\eqref{eq:reasonable:potential}).

Let $\hat{e}$ be the last task in $\chi(\sigma\circ e)$. Formula~\eqref{eq33}
is simply the local cost for algorithm $\hat{A}$ on task $\hat{e}$. Thus, we
have bounded the cost for algorithm $A$ on task $e$ to be no more than the
cost for algorithm $\hat{A}$ on task $\hat{e}$.
\end{proof}

\medskip
\emph{Proof of Theorem \ref{th:Aisrhat}.} We associate a weight vector
$\alpha$ and a bounded potential function $\Phi$ with algorithm $A$, where
\begin{align*}
\alpha(v)&=\hat{\alpha}(z_\ell) \alpha_\ell(v)\quad\text{for } v\in
 M_\ell\,; &
 \Phi(w) &=\hat{\Phi}(\hat{w}) + r \sum_i  \hat{\alpha}(z_i)
 \Phi_i(w_i)/\hat{r}_i .
\end{align*}
We remark that from \lemref{hatw} and \lemref{welleqw} it follows that
$\hat{w}$ and $w_i$ are determined by $w$, so $\Phi(w)$ is well defined.

We derive the following upper bound on the cost of $A$:
\begin{align}
  \cost_A&(\sigma\circ e) - \cost_A(\sigma) \nonumber \\
  & \leq \cost_{\hat{A}}(\chi(\sigma\circ e))- \cost_{\hat{A}} (\chi(\sigma))\label{eq40}  \\
 &\leq r \Bigl (\sum_i \hat{\alpha}(z_i) \hat{w}^e(z_i) - \sum_i \hat{\alpha}(z_i)
\hat{w}(z_i)\Bigr )  -\Bigl(\hat{\Phi}(\hat{w}^e) -
\hat{\Phi}(\hat{w})\Bigr) \label{eq41} \displaybreak[1]\\
 &= r\biggl(\sum_i \sum_{v\in M_i}
 \hat{\alpha}(z_i)\alpha_i(v) w_i^e(v)  - \sum_i \sum_{v\in M_i}
 \hat{\alpha}(z_i)\alpha_i(v) w_i(v) \biggr) \nonumber \\
 & \quad - \Bigl(\bigl(\hat{\Phi}(\hat{w}^e) + r \sum_i \hat{\alpha}(z_i)
 \Phi_i(w_i^e)/\hat{r}_i \bigr)  - \bigl(\hat{\Phi}(\hat{w}) + r \sum_i \hat{\alpha}(z_i)
 \Phi_i(w_i)/\hat{r}_i \bigr) \Bigr) \label{eq43}  \\
 &= r( \la \alpha, w^e \ra - \la \alpha,w \ra) 
   - (\Phi(w^e)-\Phi(w)). \label{eq45}
\end{align}

Inequality \eqref{eq40} follows from Proposition~\ref{lemma:samecompratio}.
Inequality \eqref{eq41} is implied as $\hat{A}$ is a sensible $r$
competitive algorithm.
We obtain \eqref{eq43} by substituting $\hat{w}^e(z_i)$ and $\hat{w}(z_i)$
according to \lemref{hatw} and rearranging the summands.
Equation~\eqref{eq45} follows from the definition of $\alpha$ and
$\Phi$ above, and using \lemref{welleqw}.

We now  prove that $A$ is $(\beta,\eta)$-constrained. It follows from
\lemref{betatagc} and \lemref{welleqw} that the condition on $\beta$ is
satisfied (see \defref{betaetacons}). It remains to show the condition on
$\eta$:

\begin{align}
\|\Phi\|_\infty &\leq \|\hat{\Phi}\|_\infty + r\sum_i
\hat{\alpha}(z_i)\|\Phi_i\|_\infty /\hat{r}_i \label{eq51}
 \\ &\leq \hat{\eta}r\cdot \diam(\hat{M}) + r \sum_i
 \hat{\alpha}(z_i) \eta_i \hat{r}_i\cdot \diam(M_i)/\hat{r}_i \label{eq52} \\
 &\leq r\cdot \diam(M) \bigl(\hat{\eta}\tfrac{\diam(\hat{M})}{\diam(M)} +
 \max_i \{\eta_i\,\tfrac{ \diam(M_i)}{\diam(M)}\}\bigr) \nonumber  \\
  &=   r\cdot\diam(M)\eta \nonumber.
\end{align}
Inequality \eqref{eq51} follows by the definition of $\Phi$, \eqref{eq52}
follows because $\hat{A}$ is $(\hat{\beta},\hat{\eta})$-constrained and $A_i$
is $(\beta_i, \eta_i)$-constrained, $1\leq i \leq b$.

We have therefore shown that $A$ is a $(\beta,\eta)$-constrained and
$r$-competitive algorithm. \endproof

\subsection{Constrained Algorithms} \label{sec:constrained}

Theorem~\ref{th:Aisrhat} assumes the existence of constrained
algorithms. In this section we show how to obtain such algorithms.
The proof is motivated by similar ideas from \cite{Sei99,BBBT97}.

\begin{definition}\label{def:variant}
Fix a metric space $M$ on $b$ states and cost ratios
$r_1,\ldots,r_b$. Assume that for all $s>0$ there is a
$(\beta,\eta)$ constrained $f(s)$ competitive algorithm $A_s$ for
the UMTS $U_s=(M;r_1,\ldots,r_b;s)$ against reasonable task
sequences. For $\rho>0$ we define the $\rho$-variant of $A_s$ (if
it exists) to be a $(\beta\rho,\eta\rho)$ constrained $f(s/\rho)$
competitive algorithm for $U_s$.
\end{definition}

\begin{lemma} \lemlab{algrhom}
Let $0<\beta \leq 1$ and $0<\beta/\rho \leq 1$. Assume there
exists a $(\beta/\rho,\eta/\rho)$-constrained and $r$-competitive
online algorithm $A'$ for the UMTS $U'=(\rho M; r_1, \ldots, r_b;
s/\rho )$. Then there exists a $(\beta, \eta)$-constrained and $r$
competitive algorithm $A$ for the UMTS $U=(M; r_1, \ldots,
r_b;s)$.
\end{lemma}
\begin{proof}
Algorithm $A$ on the UMTS $U$ simulates algorithm $A'$ on the UMTS $U'$ by
translating every task $(v,\delta)$ to task $(v',\delta)$. The probability
that $A$ associates with state  $v$ is the same as the probability that
algorithm $A'$ associates with state $v'$. If the task sequence for $A'$ is
reasonable then the simulated task sequence for $A'$ is also reasonable
simply because the probabilities for $v$ and $v'$ are identical.

The costs of $A$ or $A'$ on task $(v,\delta)$ or $(v',\delta)$ can
be partitioned into moving costs and local costs. As the
probability distributions are identical, the local costs for $A$
and $A'$ are the same. The unweighted moving costs for $A$ are
$1/\rho$ the unweighted moving costs for $A'$ because all
distances are multiplied by $1/\rho$. However, the moving costs
for $A'$ are the unweighted moving costs multiplied by a factor of
$s/\rho$ whereas the moving costs for $A$ are the unweighted
moving costs multiplied by a factor of $s$. Thus, the moving costs
are also equal.

To show that $A$ is $(\beta,\eta)$-constrained (and hence reasonable) we
first need to show that if the work functions in $U$ and $U'$ are equal,
then this implies that if $u$ and $v$ are two states such that $w(u)\geq
w(v)+\beta \dist_M(u,v)$ then $p(u)=0$. This is true because $A'$ is
$(\beta/\rho, \eta/\rho)$-constrained, and thus $w(u') \geq w(v') +
(\beta/\rho)\cdot \dist_{\rho M}(u',v')$ implies a probability of zero on
$u'$ for $A'$ which implies a probability of zero on $u$ for $A$. Next, one
needs to show that the work functions are the same, this can be done using an
argument similar to the proof of \lemref{welleqw}.

As the work functions and costs are the same for the online algorithms $A$
and $A'$ it follows that we can use the same potential function. To show that
$|\Phi| \leq \eta\cdot \diam(M)$ we note that $|\Phi|\leq (\eta/\rho)
\diam(\rho M)$.
\end{proof}

\begin{observation}
\label{obs:natural} Assume there exists a
$(\beta,\eta)$-constrained and $r$-competitive algorithm $A$ for a
UMTS $U=(M; r_1,\ldots,r_b ; s)$.
 Then, for all $\rho>0$, a natural modification
of $A$, $A'$, is a $(\beta,\eta)$-constrained, $r$-competitive
algorithm for the UMTS $U'=(\rho M; r_1, \ldots, r_b ;s)$.
\end{observation}

\begin{lemma}\label{lemma:variant}
Under the assumptions of Definition~\ref{def:variant}, for all $\rho>0$ such
that $\beta \rho \leq 1$, and for all $s>0$, the $\rho$-variant of $A_s$
exists.
\end{lemma}
\begin{proof}
For all $\rho>0$ such that $\beta\rho\leq 1$:
 \begin{enumerate}
 \item  By the assumption, there exists a $(\beta,\eta)$-constrained,
$f(s/\rho)$-competitive algorithm for the UMTS $(M; r_1,\ldots,r_b; s/\rho)$.
 \item It follows from \lemref{algrhom} that there exists an online
 algorithm that is $(\rho\beta,\rho\eta)$-constrained, $f(s/\rho)$-competitive for the UMTS
 $(\rho^{-1}M; r_1, \ldots, r_b; s)$.
 \item It now follows from Observation \ref{obs:natural} that there exists a
 $(\rho\beta,\rho\eta)$-constrained, $f(s/\rho)$-competitive online algorithm for the UMTS
 $(M; r_1,\ldots,r_b; s )$. This means
 that the $\rho$ variant of $A_s$ exists.
\end{enumerate}
\end{proof}

\section{The Uniform Metric Space} \label{section:building}

Let $\U_b^d$ denote the metric space on $b$ points where all
pairwise distances are $d$ (a uniform metric space). In this
section we develop algorithms for UMTSs whose underlying metric is
uniform. We begin with two special cases that were previously
studied in the literature.

The first algorithm works for the UMTS
$U=(\U_b^d;(r_1,\ldots,r_b);s)$, $b\geq 2$, and $r_1=r_2=\cdots
r_b$. However, it can be defined for arbitrary cost ratios. The
algorithm, called \textsc{OddExponent}, was defined and analyzed
in \cite{BBBT97}. Applying our terminology to the results of
\cite{BBBT97}, we obtain:

\begin{lemma}\lemlab{R1}
\textsc{OddExponent} is $(1,1)$-constrained, and $(\max_i r_i + 6 s \ln
b)$-competitive.
\end{lemma}
\begin{proof}
Algorithm \textsc{OddExponent}, when servicing a reasonable task
sequence, allocates for configuration $v$ the probability \( p(v)=
\frac{1}{b}+ \frac{1}{b} \sum_{u} \bigl( \frac{w(u) - w(v)}{d}
\bigr )^t \), where $t$ is chosen to be an odd integer in the
range $[\ln b, \ln b+2)$.

In our terminology, Bartal {\etal} \cite{BBBT97} prove that
\textsc{OddExponent} is sensible, $(\max_i r_i +6 s \ln
b)$-competitive and that the associated potential function
$|\Phi_1| \leq (\max_i r_i/(t+1) +s)d \leq (1/\lceil \ln b
\rceil)(\max _i r_i+6s \ln b)d$. This implies that
\textsc{OddExponent} is $(1,1/\lceil \ln b \rceil)$-constrained.
\end{proof}

The second algorithm works for the two point UMTS
$U=(\U_2^d;r_1,r_2;s)$. The algorithm, called \textsc{TwoStable},
was defined and analyzed in \cite{Sei99} and \cite{BBBT97}; based
on an implicit description of the algorithm that appeared
previously in~\cite{BKRS00}. Applying our terminology to the
results of \cite{Sei99,BBBT97}, we obtain:

\begin{lemma}\lemlab{R2}
\textsc{TwoStable} is $(1,4)$-constrained, and $r$ competitive where \[ r=
r_{1} + \frac{r_{1} - r_{2}}{e^{(r_{1} - r_{2})/s}-1} = r_{2} + \frac{r_{2} -
r_{1}}{e^{(r_{2} - r_{1})/s}-1}. \]
\end{lemma}
\begin{proof}
\textsc{TwoStable} works as follows: Let $y=w(v_1) - w(v_2)$, and
$z=(r_{1}-r_{2})/s$. The probability on point $v_1$ is \( p(v_1)=
\bigl(e^{z}-  e^{z(\frac{1}{2}+\frac{y}{2d} )}\bigr ) / \bigl (e^{z}-1 \bigr
). \) \textsc{TwoStable} is shown to be sensible and $r$ competitive in
\cite{BBBT97,Sei99} and the potential function associated with
\textsc{TwoStable}, $\Phi_2$, obeys $|\Phi_2|\leq (2r_{2}+s)d$.

It remains to show that $|\Phi_2|\leq 4rd$. We use the fact that,
in general, if $|z|\leq 1/2$ then $ 1/2 \leq z/(e^z -1)$, and do a
simple case analysis. If $\max\{r_1,r_2\} >\frac{1}{2} s$ then
$|\Phi_2| \leq (2 r_2 +s)d \leq (2r +2r)d \leq 4rd$. Otherwise,
$|z|\leq 1/2$, so \( r= r_2 + \frac{z}{e^{z}-1}s\geq r_2+
\frac{s}{2} \). Hence $|\Phi_2| \leq 2 r d$.
\end{proof}

To gain an insight about the competitive ratio of \textsc{TwoStable}, we
have the following proposition.

\begin{proposition} \proplab{mts-two-lnub}
Let \( f(s,r_1,r_2)= r_{1} + (r_{1} - r_{2})/ \bigl(e^{(r_{1} - r_{2})/s}-1
\bigr ).\) Let  $x_1,x_2\in \Re^+$ such that $r_1 \leq 2 s (\ln x_1+1)$ and
$r_2 \leq 2 s (\ln x_2+1)$. Then $f(s,r_1,r_2) \leq 2 s (\ln(x_1+x_2)+1)$.
\end{proposition}
\begin{proof}
First we show that $f$ is a monotonic non-decreasing function of both $r_1$
and $r_2$. Since the formula is symmetric in $r_1$ and $r_2$ it is enough to
check monotonicity in $r_1$. Let $x=(r_1- r_2)/s$, it suffices to show that
$g(x)=sx+r_2+ s x/(e^x-1)$ is monotonic in $x$. Taking the derivative
\[ g'(x)=s\cdot\frac{e^x (e^x -(1+x))}{(e^x-1)^2} \geq 0 ,
\text{ since } e^x \geq 1+x.\]

Therefore we may assume that $r_1=2 s(\ln x_1 +1)$ and $r_2=2 s(\ln x_2 +1)$.
Without loss of generality we can assume that $x_1\geq x_2$ and let $y\geq 2$
be such that $x_1=(x_1+x_2)(1- 1/y)$. By substitution we get
\begin{math}
  r_1 - r_2 = 2 s \ln (y-1)
\end{math}
and
\begin{multline*}
f(s,r_1,r_2)  = r_1+ \frac{r_1 - r_2}{e^{(r_1- r_2)/s} -1}
 = 2 s \Bigl( \ln (x_1+x_2) +1  + \ln (y-1) - \ln y + \frac{\ln (y-1)}{(y-1)^2 -1} \Bigl ) \\
 \leq 2s \Bigl (\ln (x_1+x_2) +1 -\frac{1}{y} +\frac{\ln (y-1)}{(y-1)^2 -1}
 \Bigr).
\end{multline*}
 We now prove that for $y\geq 2$,
\( -\frac{1}{y} + \frac{\ln (y-1)}{(y-1)^2 -1}\leq 0.\) When $y$ approaches
2, the limit of the expression is zero. For $y>2$, we multiply the left side
by $(y-1)^2-1$, and get $g(y)= -(y-2)+ \ln (y-1)$. Since $g(2)=0$ and
$g'(y)=-1 + 1/(y-1)<0$ for $y>2$, we are done.
\end{proof}

\begin{figure}
\centering
\epsfig{file=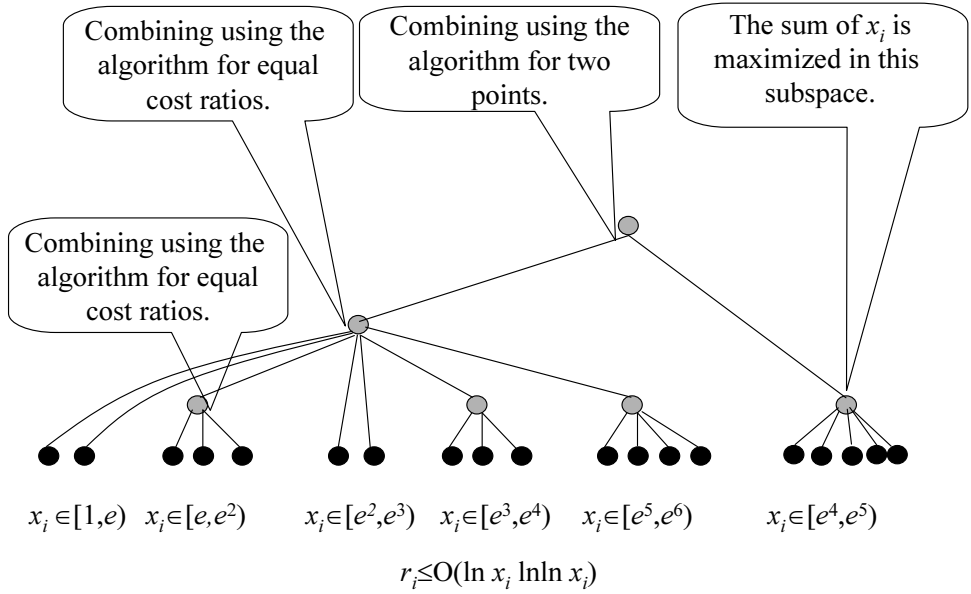, angle=-90, scale=0.6} 
 \caption{Schematic description of {\combn}.}
 \label{fig:R3}
\end{figure}

\medskip

We next describe a new algorithm, called {\combn}, defined on a
UMTS $U=(\U^d_b; r_1, \dots, r_b;s)$. This algorithm is inspired
by \textsl{Strategy~3} \cite{BBBT97}. Like \textsl{Strategy~3},
{\combn} combines \textsc{OddExponent} and \textsc{TwoStable} on
subspaces of $\U^d_b$, however, it does so in a more sophisticated
way that is impossible using the combining technique of
\cite{BBBT97}. Fig.~\ref{fig:R3} presents the scheme of the
combining process.

\begin{algrthree}
As discussed in Observation~\ref{obs:w/o-dr}, we may assume that $s=1$. Let
$x_i$ be the minimal real number such that $r_i\leq 100  \ln x_i \ln \ln
x_i$ and $x_i \geq e^{e^6+1}$, and let $x$ denote $\sum_i x_i$. For a set
$S\subset M^d_b$ let $U(S)$ denote the UMTS induced by $U$ on $S$.

Let $\U^d_b=\{v_1, \ldots, v_b\}$, where $v_i$ has cost ratio $r_i$. We
partition the points of $\U^d_b$ as follows: let \( Q_\ell = \{ v_i:\;
e^{\ell-1} \leq x_i < e^\ell \} \). Let $P = \{ Q_\ell : |Q_\ell| \geq  \ln x
\} \cup \{ \{ v \} : v\in Q_\ell \text{ and } |Q_\ell| <  \ln x \}$, $P$ is a
partition of $\U^d_b$. For $S\in P$ let $x(S) = \sum_{v_i\in S} x_i$. Without
loss of generality we assume $P=\{S_1, S_2, \ldots, S_{b'}\}$ where $b'=|P|$
and $x(S_j) \geq x(S_{j+1})$, $1\leq j \leq b'-1$.

We associate with every set $S_i$ an algorithm $A(S_i)$ on the
UMTS $U(S_i)$. If $|S_i|\geq \ln x$ we choose $A(S_i)$ to be the
$(1/10)$-variant of \textsc{OddExponent}. If $|S_i|< \ln x$ then
$|S_i|=1$ and we choose $A(S_i)$ to be the trivial algorithm on
one point, this algorithm has a competitive ratio equal to the
cost ratio, and it is $(0,0)$-constrained. Let $r(S_i)$ denote the
competitive ratio of $A(S_i)$ on $U(S_i)$.

If $b'=1$ we choose {\combn} to be $A(S_1)$ and we are done. If $b' \geq 2$,
let $\tilde{M} = \cup_{i=2}^{b'} S_i$. We want to construct an algorithm,
$A(\tilde{M})$, for $U(\tilde{M})$. If $b'=2$, we choose $A(\tilde{M})$ to
be $A(S_2)$. Otherwise, we apply Theorem~\ref{th:Aisrhat} on $\tilde{M}$
with the partition $\{S_2, \ldots, S_{b'}\}$. We define $\hat{M}$ from
Theorem \ref{th:Aisrhat} to be $\U^d_{b'-1}$. Likewise, $\hat{A}$ from
Theorem \ref{th:Aisrhat} is the application of the $(1/5)$-variant of
\textsc{OddExponent} on $\hat{U} = (\U^d_{b'-1}; r(S_2), \ldots,
r(S_{b'}))$. Let $r(\tilde{M})$ denote the competitive ratio of $\hat{A}$.

Next, we choose the  partition $\{S_1, \tilde{M}\}$ of $\U^d_b$. We combine the
two algorithms $A(S_1)$ and $A(\tilde{M})$ using the $(1/10)$ variant of
\textsc{TwoStable} (this is the $\hat{A}$ required in Theorem
\ref{th:Aisrhat}) on the UMTS $(\U^d_2;r(S_1),r(\tilde{M}))$ (the UMTS
$\hat{U}$ of Theorem \ref{th:Aisrhat}). We denote the competitive ratio of
$\hat{A}$ by $r$. The resulting combined algorithm, $A(M)$, is our final algorithm,
{\combn}.
\end{algrthree}

\begin{lemma}\label{lemma:R3}
Given that $x=\sum_i x_i$,  $r_i \leq 100 s \ln x_i \ln \ln x_i$, and $x_i
\geq  e^{e^6+1}$, algorithm {\combn} for the UMTS $U=(\U^d_b;r_1,$
$\ldots,r_b;s)$ is $(1,1/2)$-constrained and $r$-competitive, where  $r\leq
100 s \ln x \ln\ln x$.
\end{lemma}
\begin{proof}
As before, without loss of generality, we assume $s=1$. First we calculate
the constraints of the algorithm.

From \lemref{R1} and Lemma~\ref{lemma:variant}, $A(S_i)$ is
$(1/10,1/10)$-const\-rained, for every $1\leq i \leq b'$. We would
like to show that $A(\tilde{M})$ is $(1/2,3/10)$-constrained. If
$b'=2$ then it obviously $(1/10,1/10)$-constrained. Otherwise,
($b'>2$), the combining algorithm for $\tilde{M}$ is the
$(1/5)$-variant of \textsc{OddExponent} which is
$(1/5,1/5)$-constrained. Hence, from \eqref{eqbeta2}, $\beta \leq
1/5+1/10+1/10+1/10 = 1/2$, and from \eqref{eqeta2}, $\eta \leq
1/5+1/10 = 3/10$. From Corollary~\ref{cor:Aisrhat}, $A(\tilde{M})$
is $r(\tilde{M})$ competitive.

The $(\beta,\eta)$-constraints  of algorithm {\combn} are calculated as
follows: The $(1/10)$-variant of \textsc{TwoStable} is $(1/10,2/10)$
constrained, therefore $\beta= 1/10+1/10+1/2+3/10=1$ and
$\eta=2/10+3/10 = 1/2$. From Corollary~\ref{cor:Aisrhat}, $A(M)$ is
$r$-competitive.

To summarize, {\combn} is $(1,1/2)$-constrained and $r$-compe\-titive
algorithm for the UMTS $U$.

\newcommand{\coffa}{100}

It remains to prove the bound on $r$. First we show that $r(S_j)\leq 100 s
\ln x(S_j) \ln\ln x(S_j)$ for all $1\leq j \leq b'$. If $|S_j|=1$, we are
done. Otherwise, $|S_j| \geq \ln x$, and $S_j=Q_\ell$ for some $\ell$.
\begin{align}
r(S_j)
  & \leq  \coffa  \ln e^\ell \ln \ln e^\ell + 6\cdot10  \ln |S_j|
\label{eq61}\\
       & \leq  \coffa  \bigl(\ln e^{\ell-1} \ln \ln e^{\ell-1} + \ln \ell +
       \tfrac{1}{\ell-1}  \ln e^{\ell-1}  \bigr) + 60  \ln |S_j| \nonumber \\
       &\leq  \coffa  \bigl (\ln e^{\ell-1} \ln \ln e^{\ell-1}
             + \ln \ln x + \tfrac{60}{\coffa}\ln |S_j| + 1 \bigr)\label{eq62}
             \\
       &\leq  \coffa  \bigl( \ln e^{\ell-1} \ln \ln e^{\ell-1} + 2 \ln |S_j| \bigr )\label{eq63} \\
       &\leq \coffa  \ln (|S_j| e^{\ell-1}) \ln \ln (|S_j| e^{\ell-1})
       \nonumber \\
       &\leq  \coffa \ln x(S_j) \ln \ln x(S_j). \label{eq65}
\end{align}

Inequality~\eqref{eq61} is derived as follows.
Since $S_j = Q_\ell$, it follows that  $r_i \leq 100 s
\ln e^\ell \ln\ln e^\ell$ for all $v_i\in S_j$. By the bound on the
competitive ratio of the $(1/10)$ variant of \textsc{OddExponent} (See
\lemref{R1} and Lemma~\ref{lemma:variant}) we obtain \eqref{eq61}.
Inequality~\eqref{eq62} follows since $\ell\leq \ln x$. Inequality
\eqref{eq63} follows because $\ln |S_j| \geq \ln \ln x$, and $\ln \ln x \geq
6$. The last inequality follows because $e^{\ell-1}$ is a lower bound on
$x_i$ for $v_i \in S_j$ and thus $|S_j|e^{\ell-1}\leq x(S_j)$.

Observe that $b'\leq \ln^2 x$ as there are at most $\ln x$ sets $Q_i$, and
each such set contributes at most $\ln x$ sets $S_i$ to $P$.  We next derive a
bound on $r(\tilde{M})$.
\begin{eqnarray}
r(\tilde{M}) &\leq&\max_{2\leq i \leq b'} r(S_i) + 6\cdot 5 \cdot
\ln(b'-1) \label{eq71} \\
 &\leq& 100  \cdot \ln x(S_2) \ln\ln x + 30 \cdot (2 \ln \ln x ) \label{eq72} \\
 &=& 100 (\ln x(S_2) + 0.6) \ln \ln x. \nonumber
\end{eqnarray}
Inequality \eqref{eq71} follows since the algorithm used is a $(1/5)$ variant
of \textsc{OddExponent}. Inequality \eqref{eq72} follows by using the
previously derived bound on $r(S_i)$ and noting  that $x(S_2)$ is maximal
amongst $x(S_2),\ldots, x(S_{b'})$ and that $x(S_i)\leq x$.

 From Lemma~\ref{lemma:variant} we know that the competitive ratio of the
$(1/10)$-variant of \textsc{TwoStable} is $f(10 , r(S_1), r(\tilde{M}))$
where $f$ is the function as given in \propref{mts-two-lnub}. We give an
upper bound on $f(10 , r(S_1), r(\tilde{M}))$ using \propref{mts-two-lnub}.
To do this we need to find values $y_1$ and $y_2$ such that
\begin{align*}
r(S_1) \leq 100  \ln x(S_1) \ln \ln x &= 2\cdot 10  (\ln y_1 +1) \\
r(\tilde{M}) \leq 100  (\ln x(\tilde{M}) +0.6) \ln \ln x &= 2 \cdot 10  (\ln y_2
+1).
\end{align*}

Indeed, the following values satisfy the conditions above: $y_1=x(S_1)^{5
\ln \ln x}/e$ and $y_2=(e^{0.6}x(\tilde{M}))^{5 \ln \ln x}/e$. Using
\propref{mts-two-lnub} we get a bound on $r$ as follows
\begin{align}
 r
  &\leq 2\cdot 10  (\ln (y_1 + y_2  ) +1)  \label{eq81}\\
 &\leq  20  \ln \bigl( x(S_1)^{5 \ln \ln x} +
       (e^{0.6} x(\tilde{M}))^{5 \ln \ln x} \bigr) \nonumber \\
  & \leq  20  \ln \bigl( x(S_1)^{5 \ln \ln x} +
  (2^{5 \ln \ln x}-1) x(\tilde{M})^{5 \ln \ln x}
  \bigr ) \label{eq82}\\
  & \leq  20   \ln \bigl ( (x(S_1)+x(\tilde{M}))^{5 \ln \ln x}
  \bigr)\label{eq83}   \\
   &\leq \coffa  \ln x \ln \ln x. \nonumber
\end{align}

Inequality \eqref{eq81} follows from \propref{mts-two-lnub}.
Inequality~\eqref{eq82} follows because $\ln \ln x \geq 6$. Inequality
\eqref{eq83} follows since, in general, for $a\geq b>0$ and $z\geq 1$,
 \( a^z + (2^z-1)b^z \leq(a+b)^z \). This is because
for $a=b$ it is an equality, and the derivative with respect to
$a$ of the RHS is clearly larger than the derivative with respect
to $a$ of the LHS.
\end{proof}

Next, we present a better algorithm when all the cost ratios but one are
equal.

\begin{lemma} \label{lemma:R4}
Given a UMTS $U=(\U^d_b; r_1, r_2, \ldots, r_b)$ with $r_2=r_3 =
\cdots = r_b$, there exists a $(1,3/5)$-constrained and
$r$-competitive online algorithm, {\wcombn}, where \[ r = 30
\Bigl( \ln \bigl( e^{\frac{r_1}{30} -\frac{1}{3}} + (b-1)
e^{\frac{r_2}{30} - \frac{1}{3}} \bigr) + \tfrac{1}{3}\Bigr). \]
\end{lemma}
\begin{proof} The proof is a simplified version of the proof of
Lemma~\ref{lemma:R3}, and we only sketch it here. We define $x_1$, $x_2$,
such that
\begin{align*}
 r_1 &=30  (\ln x_1+\tfrac{1}{3}) = 2\cdot5 \cdot  (\ln x_1^3 +1),  &
 r_2 &=30  (\ln x_2 +\tfrac{1}{3})= 2\cdot 5 \cdot  (\ln x_2^3 +1).
\end{align*}
Let $\tilde{M}=\{v_2,\ldots v_b\}$. We use a $(1/5)$ variant of
\textsc{OddExponent} on the UMTS $U(\tilde{M})$. The competitive ratio of
this algorithm is at most
\begin{equation*}
r(\tilde{M}) \leq r_2 + 30  \ln (b-1)  \leq 30
\bigl(\ln((b-1)x_2)+\tfrac{1}{3} \bigr) =  10 \bigl ( \ln ((b-1)x_2)^3 +1
\bigr )
\end{equation*}
and it is $(1/5,1/5)$ constrained. We combine it with the trivial algorithm
for $U(\{v_1\})$ using a $(1/5)$ variant of algorithm \textsc{TwoStable},
the resulting algorithm is $(1,3/5)$ constrained, and by
\propref{mts-two-lnub} we have
\begin{multline*}
r\leq 10  (\ln(x_1^3 +((b-1)x_2)^3 +1)
 \leq  10  ( \ln (x_1+(b-1)x_2)^3 +1)
 =30 ( \ln  (x_1+(b-1)x_2) + \tfrac{1}{3} ).
\end{multline*}
Substituting for $x_i$ gives the required bound.
\end{proof}

\section{Applications} \seclab{app}

\subsection{An {\large ${O((\log n \log\log n)^2)}$} Competitive algorithm for MTSs}
\seclab{anymts-ub}

Bartal~\cite{Bar96} defines a class of decomposable spaces called
\emph{hierarchically well separated trees} (HST).\footnote{The definition
given here for $k$-HST differs slightly from the original definition given in
\cite{Bar96}. We choose the definition given here for simplicity of the
presentation. For $k >1$ the metric spaces given by these two definitions
approximate each other to within a factor of $k/(k-1)$.}

\begin{definition} \deflab{hst}
For $k\geq 1$, a $k$-\emph{hierarchically well-separated tree}
($k$-HST) is a metric space defined on the leaves of a rooted tree
$T$. Associated with each vertex $u\in T$  is a real valued label
$\Delta(u) \ge 0$, and $\Delta(u)=0$ if and only if $u$ is a leaf
of $T$. The labels obey the rule that for every vertex $v$, a
child of $u$, $\Delta(v)\leq \Delta(u)/k$. The distance between
two leaves $x,y\in T$ is defined as $\Delta(\lca(x,y))$, where
$\lca(x,y)$ is the least common ancestor of $x$ and $y$ in $T$.
Clearly, this is a metric.
\end{definition}

Bartal~\cite{Bar96,Bar98} shows how to approximate any  metric
space using an efficiently constructible probability distribution
over a set of $k$-{\hst}s . His result allows to reduce a MTS
problem on an arbitrary metric space to MTS problems on HSTs.
Formally, he proves the following theorem.

\begin{theorem}[\cite{Bar98}]
\theolab{bar98} Suppose there is a $r$-competitive  algorithm for any
$n$-point $k$-{\hst} metric space. Then there exists an $O(r k \log n \log
\log n)$-competitive randomized algorithm for \emph{any} $n$-point metric
space.
\end{theorem}

Thus, it is sufficient to construct an online algorithm for a metrical task system where the underlying
metric space is a $k$-{\hst}. Following \cite{BBBT97} we use the unfair MTS
model to obtain an online algorithm for a MTS over a $k$-{\hst} metric space.

\begin{RHST}
We define the algorithm {\rhst}($T$) on the metric space $M(T)$, where $T$
is a $k$-HST with $k\geq 5$. Algorithm {\rhst}($T$) is defined inductively on
the size of the underlying HST, $T$.

When $|M(T)|=1$, {\rhst}($T$) serves all task sequences optimally. It is
$(0,0)$-constrained. Otherwise, let the children of the root of $T$ be $v_1,
\ldots, v_b$, and let $T_i$ be the subtree rooted at $v_i$. Denote
$d=\Delta(T)$, and so  $\diam(T_i) \leq d/k$. Every algorithm {\rhst}($T_i$)
is an algorithm for the UMTS $U_i=(M(T_i);1,\ldots,1;1)$.

We construct a metric space $\hat{M}=\U^d_b$, and define cost
ratios $r_1, \ldots, r_b$ where $r_i = r(T_i)$ is the competitive
ratio of {\rhst}($T_i$). We now use Theorem~\ref{th:Aisrhat} to
combine algorithms {\rhst}($T_i$). The role of $\hat{A}$ is played
by the $(1/2)$ variant of {\combn} on the unfair metrical task
system $\hat{U}=(\hat{M};r_1, \ldots,r_b;1)$. The combined
algorithm is a {\rhst}($T$) on the UMTS $(M(T);1,\ldots,1;1)$.

We remark that the application of Theorem \ref{th:Aisrhat} requires that the
algorithms will be constrained. We show that this is true in the following lemma.
\end{RHST}

\begin{lemma} \label{lemma:RHST}
The algorithm {\rhst}($T$)  is $O(\ln n \ln\ln n)$,
where $n=|M(T)|$.
\end{lemma}
\begin{proof}
Let $n'=e^{e^6+1} n$. We prove by induction on the depth of the tree that
{\rhst}($T$) is $(1,1)$-constrained and $200 \ln n' \ln \ln n'$-competitive.

When $|M(T)|=1$, it is obvious. Otherwise, let $n_i = |M(T_i)|$,
$n'_i=e^{e^6+1} n_i$, and $n'=\sum_i n'_i$. We assume inductively that each
of the {\rhst}($T_i$) algorithms is $(1,1)$-constrained and \linebreak[3]
$200 \ln n'_i \ln \ln  n'_i$ competitive on $M(T_i)$. The combined algorithm,
{\rhst}($T$), is $(\beta, \eta)$-constrained. From \eqref{eqbeta2}, and
given that $k\geq 5$, we get that
\begin{equation*}
\beta \leq  \max \{1, \tfrac{1}{2} + \tfrac{1}{k} +\tfrac{1}{k}
 + \tfrac{1}{2k} \}
 \leq  \max \{1,1\} = 1.
 \end{equation*}
 From \eqref{eqeta2} we obtain that $\eta \leq \tfrac{1}{2}  +
\tfrac{1}{k} \leq 1$, for $k\geq 5$. This proves that the algorithm is well
defined and $(1,1)$ constrained.

We next bound the competitive ratio using Lemma \ref{lemma:R3}.
Lemma~\ref{lemma:variant} implies that the competitive ratio
obtained by the $(1/2)$ variant of {\combn} on $(\hat{M};r_1,
\ldots, r_b)$ is the same as the competitive ratio attained by
{\combn} on $(\hat{M};r_1,\ldots,r_b;2)$. The values $(x_i)_i$
computed by {\combn} are at most $(n'_i)_i$, respectively. Hence
it follows from Lemma~\ref{lemma:R3} that the competitive ratio of
{\rhst}($T$) is at most $100\cdot 2 \ln x \ln \ln x \leq 200 \ln
n' \ln \ln  n'$, since $x=\sum_i x_i$.
\end{proof}

Since every HST $T$ can be $5$-approximated by a 5-HST $T'$ (see
\cite{Bar98}), the bound we have just proved holds for any HST.

Combining \theoref{bar98} with Lemma \ref{lemma:RHST}, it follows that
\begin{theorem}
For any MTS over an $n$-point metric space, the randomized competitive ratio
is $O((\log n \log\log n)^2)$.
\end{theorem}

\subsection{{$K$}-Weighted Caching on {$K+1$} Points}

Weighted caching is a generalized paging problem where there is a different
cost to fetch different pages. This problem is equivalent to the $K$-server
problem on a \emph{star metric space} \cite{You91A,BFT96}. A star metric
space is derived from a depth one tree with distances on the edges, the
points of the metric space are the leaves of the tree and the distance
between a pair of points is the length of the (2 edge) path between them.
This is so, since we can assign any edge $(r,u)$ in the tree a weight of
half the fetch cost of $u$. Together, an entrance of a server into a leaf
from the star's middle-point (page in)  and leaving the leaf  to the star's
middle point (page out) have the same cost of fetching the page.

The $K$-server problem on a metric space of $K+1$ points is a special case of
the metrical task system problem on the same metric space, and hence any
upper bound for the metrical task system translates to an upper bound for the
corresponding $K$-server problem.

Given a star metric space $M$, we $12$-approximates it with a $6$-HST $T$.
$T$ has the special structure that for every internal vertex, all children
except perhaps one, are leaves. It is not hard to see  that one can find such
a tree $T$ such that for any $u,v\in M$, \( \dist_M(u,v) \leq \dist_{T}(u,v)
\leq 12 \cdot \dist_{M}(u,v).\) Essentially, the vertices furthest away from
the root (up to a factor of 6) in the star are children of the root of $T$
and the last child of the root is a recursive construction for the rest of
the points.

We now follow the construction of {\rhst} given in the previous section, on
an 6-HST $T$, except that we make use of $(1/2)$-variant of {\wcombn} rather
than $(1/2)$-variant of {\combn}. The special structure of $T$ implies that
all the children of an inner vertex, except perhaps one, are leaves and
therefore have a trivial $1$-competitive algorithm on their ``subspaces".
Hence we can apply {\wcombn}. Using Lemma~\ref{lemma:R4} with
induction on the depth of the tree,
it is easy to bound the competitive ratio on $K+1$
leaves tree to be at most $60 (\ln(K +1) +1/3)$.

Combining the above with the lower bound of \cite{BFT96} we obtain:
\begin{theorem}
The competitive ratio for the $K$-weighted caching problem on $K+1$ points is
$\Theta(\log K)$.
\end{theorem}

\subsection{A MTS on Equally Spaced Points on the Line}

The metric space of $n$ equally spaced points on the line is considered
important because of its simplicity, and the practical significance of the
$k$-server on the line (for which this problem is a special case). The best
lower bound currently known on the competitive ratio is $\Omega(\log n /
\log \log n)$ \cite{BKRS00}. Previously, the best upper bound known was
$O(\log ^3 n / \log \log n)$ due to~\cite{BBBT97}.

We are able to slightly improves the upper bound on the competitive ratio
from \secref{anymts-ub} to $O(\log^2 n)$. Bartal~\cite{Bar96} proves that
$n$ equally spaced points on the line can be $O(\log n)$ probabilistically
embedded into a set of \emph{binary} $4$-HSTs.  We present an $O(\log n)$
competitive randomized algorithm for binary $4$-HST, similar to {\rhst}
except that we make use of $(1/4)$-variant of \textsc{TwoStable} instead of
$(1/2)$-variant of {\combn}. Similar arguments show that this algorithm is
$(1,1)$-constrained, and using \propref{mts-two-lnub} we conclude that the
algorithm is $8 \ln n$ competitive. Combining the probabilistic embedding into
binary $4$-HST with the algorithm for binary $4$-HST we obtain

\begin{theorem}
The competitive ratio of the MTS problem on metric space of $n$ equally
spaced points on the line is $O(\log^2 n)$.
\end{theorem}

\section{Concluding Remarks} \seclab{concluding}

This paper present algorithms for MTS problem and related problems
with significantly improved competitive ratios. An obvious avenue
of research is to further improve the upper bound on the
competitive ratio for the MTS problem. A slight improvement to the
competitive ratio of the algorithm for arbitrary $n$-point metric
spaces is reported in~\cite{BM03}. The resulting competitive ratio
there is $O(\log^2 n \log \log n \log \log \log n)$ and the
improvement is achieved by refining the reduction from arbitrary
metric spaces to HST spaces ({\ie}, that improvement is orthogonal
to the improvement presented in this paper). However, in order to
break the $O(\log^2 n)$ bound, it seems that one needs to deviate
from the black box usage of \theoref{bar98}. Maybe the easiest
special case to start with is the metric space of equally spaced
points on the line.

Another interesting line of research would be an attempt to apply the techniques
of this and previous papers to the randomized $k$-server problem, or even for
a special case such as the randomized weighted caching on $k$ pages problem;
see also \cite{BBK99,Seiden00}.

\medskip

\paragraph{Acknowledgments} We would like to thank Yair Bartal, Avrim Blum
and Steve Seiden for helpful discussions.

\bibliographystyle{siam}
\bibliography{rmts}

\end{document}